\documentclass[aps,prb,twocolumn,superscriptaddress,unsortedaddress]{revtex4}
\usepackage{epsfig}
\usepackage{amsmath}

\newcommand{\be}{\begin{equation}}
\newcommand{\ee}{\end{equation}}
\newcommand{\bk}{{{\bf{k}}}}

\newcommand{\br}{{{\bf{r}}}}

\newcommand{\bea}{\begin{eqnarray}}
\newcommand{\eea}{\end{eqnarray}}
\newcommand{\ra}{\rangle}
\newcommand{\la}{\langle}

\newcommand{\bS}{{\bf S}}

\newcommand{\dg}{{\dagger}}
\newcommand{\pdg}{{\phantom\dagger}}
\newcommand{\nn}{\nonumber}
\newcommand{\hb}{\hat{b}}
\newcommand{\ha}{\hat{a}}

\begin{document}

\title{Quantum paramagnetic ground states on the honeycomb lattice and field-induced transition to N\'eel order}
\author{R. Ganesh}
\affiliation{Department of Physics, University of Toronto, Toronto, Ontario M5S 1A7, Canada}
\author{D. N. Sheng}
\affiliation{Department of Physics and Astronomy, California State University, Northridge, California 91330, USA}
\author{Young-June Kim}
\affiliation{Department of Physics, University of Toronto, Toronto, Ontario M5S 1A7, Canada}
\author{A. Paramekanti}
\affiliation{Department of Physics, University of Toronto, Toronto, Ontario M5S 1A7, Canada}
\affiliation{Canadian Institute for Advanced Research, Toronto, Ontario, M5G 1Z8, Canada}

\date{\today}

\begin{abstract}
{
Bi$_3$Mn$_4$O$_{12}$(NO$_3$) is a recently synthesized spin-3/2 bilayer honeycomb antiferromagnet which behaves as a spin liquid down to very low temperatures. Beyond a magnetic field of about 5T, it develops long range Neel order. Motivated by this observation, we have studied spin-$S$ Heisenberg models with next neighbor frustrating interactions as well as bilayer couplings on the honeycomb lattice.
For a model with frustrating second-neighbor exchange, $J_2$, we use a Lindemann-like criterion within spin
wave theory to show that N\'eel order melts beyond a critical $J_2$. The critical $J_2$ is found to increase in the 
presence of a magnetic field, implying the existence of a field-induced paramagnet-N\'eel transition over a range of parameters.
For the bilayer model, we use a spin-$S$ generalization of bond operator mean field theory to show that there is
a  N\'eel-dimer transition for various spin values with increasing bilayer coupling, and that the resulting interlayer
dimer state undergoes a field induced transition into a state with transverse N\'eel order.
Motivated by a broader interest in such paramagnets, we have also studied a spin-3/2 model which
interpolates between the nearest neighbor Heisenberg model and the Affleck-Kennedy-Lieb-Tasaki (AKLT) 
parent Hamiltonian. Using exact diagonalization, we have found that there
is a single Neel-AKLT quantum phase transition in this model. Computing the fidelity susceptibility and assuming a transition in the O(3) universality class, we have located the critical point of this model. In addition, we have obtained the spin gap of the AKLT parent Hamiltonian. Our numerics indicate that the AKLT state also undergoes a field induced Neel ordering 
transition. We discuss implications of some of our results for experiments on Bi$_3$Mn$_4$O$_{12}$(NO$_3$), and for numerics on the honeycomb lattice Hubbard model.}
\end{abstract}

\maketitle

The 
interplay of quantum mechanics and frustrated interactions 
in quantum magnets leads to a variety of remarkable 
phases including spin liquid Mott insulators, valence bond crystals, and Bose-Einstein condensates
of magnons.\cite{reviews} Recently, there has been tremendous interest in novel paramagnetic ground states 
on the honeycomb lattice. 
A quantum Monte Carlo study of the 
repulsive electronic Hubbard model on the honeycomb lattice
has uncovered a spin liquid ground state
\cite{assaad}
leading to a flurry of studies of honeycomb lattice
spin liquids.\cite{fawang2010,yran2010,clark2010}
Another interesting honeycomb lattice paramagnet is the $S=3/2$ Affleck-Kennedy-Lieb-Tasaki 
(AKLT) state.\cite{AKLT1987,arovas1988,KLT1988} Such an AKLT state is most easily understood
by viewing each spin-$3/2$ as being composed of three spin-1/2
moments symmetrized on-site, with each spin-1/2 moment forming a singlet with
one neighbor, leading to a ground state which respects all lattice symmetries.
This state has been suggested as an entanglement resource for universal quantum
computation.\cite{tcwei} 
Furthermore, an optical analogue of the one-dimensional AKLT state
has been realized in recent experiments,\cite{akltoptical} raising hopes for alternative realizations of
AKLT states in higher dimensions.

Interest in honeycomb lattice quantum paramagnets also stems from
experiments \cite{BiMnO} on Bi$_3$Mn$_4$O$_{12}$(NO$_3$).
The crystal field of the MnO$_6$ octahedra, together with strong Hund's coupling, leads
to Heisenberg-like spin-3/2 moments on the
Mn$^{4+}$ ions which form a bilayer honeycomb lattice. Despite the bipartite structure, and a large
antiferromagnetic Curie-Weiss constant $\Theta_{CW}\approx -257K$, this system shows no
magnetic order \cite{BiMnO} (or any other phase transition)
down to $T\!\sim\! 1 \mathrm{K}$. 
This observation hints at
frustrating interactions which may
lead to interesting paramagnetic ground states.\cite{fouet2001,mattsson1994,takano2006,mulder2010,kawamura2010,jafari2010}
Neutron scattering experiments \cite{unpub} on powder samples  of Bi$_3$Mn$_4$O$_{12}$(NO$_3$) 
in zero magnetic field indicate that there are short range spin correlations in this material,
with some antiferromagnetic coupling between the two layers forming the bilayer, but negligible
interactions between adjacent bilayers. 
Remarkably, applying a critical magnetic field, $B_c \sim 6 $ Tesla, leads to sharp
Bragg spots consistent with
three dimensional (3D) N\'eel order.\cite{unpub} 

Motivated by this broad interest in honeycomb lattice quantum paramagnets, we study
various Heisenberg models with additional exchange interactions chosen to frustrate N\'eel order.
We also consider the effect of a magnetic field on the paramagnetic states which result from the
destruction of N\'eel order. We show that applying a critical magnetic field to these paramagnetic
ground states leads to a transition into a state with long range
N\'eel order in the plane transverse to the applied field, which allows us to make connections with
ongoing experiments and predictions for numerical studies of such paramagnetic states.

We begin with a study of a model with nearest-neighbor ($J_1$) and
frustrating second-neighbor ($J_2$) exchange interactions. Such a model is relevant to 
Bi$_3$Mn$_4$O$_{12}$(NO$_3$)  as well as numerical studies of the honeycomb lattice
Hubbard model. At the classical level, 
it is known that such frustration leads to the
N\'eel state becoming unstable for $J_2/J_1 > 1/6$. In a quantum model with finite $S$, N\'eel 
order is likely to melt for smaller $J_2/J_1$, although the nature of the resulting paramagnetic ground state
is not known. Sidestepping the issue of
what state results from quantum melting,
we study the magnetic field dependence of the critical $J_2/J_1$ required to destroy the N\'eel order.
Using spin-wave theory, we show that a nonzero magnetic field enhances the critical $J_2/J_1$,
opening up a regime where applying a critical
field to the non-N\'eel state yields long-range N\'eel order.

Next, motivated by the fact that Bi$_3$Mn$_4$O$_{12}$(NO$_3$) consists of stacked
bilayers, we study a bilayer honeycomb magnet where the interlayer exchange interaction competes with
the intralayer coupling. Using a spin-$S$ generalization \cite{brijesh} of the bond operator formalism,
\cite{sachdevbhatt1990} we show that a sufficiently strong 
bilayer coupling leads to an interlayer VBS state. We obtain the N\'eel to interlayer valence-bond solid (VBS) transition point for
various spin values, which could be tested using quantum Monte Carlo numerics, as well as the triplon
dispersion in the interlayer VBS. 
We show that the presence of a magnetic field strong enough to overcome the spin
gap, results in the interlayer VBS undergoing a Bose condensation transition into a state with long range N\'eel
order in the plane transverse to the applied field.
For the spin-1/2 case, we find that the transition to the interlayer VBS state
occurs when the interlayer exchange is of the order of the interplane exchange ($J_c\sim1.3J_1$), suggesting that
spin-1/2 bilayer honeycomb magnets might be a promising system to realize this VBS state. 
For $S = 3/2$, we find that the VBS state
is only realized at large interlayer couplings, $J_c/J_1 \gtrsim 6.6$.

Recent attempts to determine the relevant
exchange couplings in  Bi$_3$Mn$_4$O$_{12}$(NO$_3$)
indicate the presence of longer range couplings in the honeycomb plane. \cite{Wadati,vandenBrink}
Furthermore, while Ref.~\onlinecite{Wadati} focused on a single layer, the {\it ab initio}
results of Ref.~\onlinecite{vandenBrink} provide evidence for large
interlayer couplings. In the light of these
reports, our work on the $J_1$-$J_2$ model and the bilayer model is
perhaps relevant to the physics of Bi$_3$Mn$_4$O$_{12}$(NO$_3$). Specifically, we find that even strong
interlayer couplings $J_c \sim 2 J_1$ as suggested by Ref.~\onlinecite{vandenBrink} cannot destabilize N\'eel order in this material.
We also note that Refs.\onlinecite{Wadati, vandenBrink} disagree on the sign of the further neighbor
couplings within the honeycomb layer, suggesting the need for further work on this issue.

Finally, from the viewpoint of broader theoretical interest,
we explore a generalized spin-3/2 model including biquadratic and bicubic spin interactions which
interpolates between a Heisenberg model and the parent Hamiltonian of the $S=3/2$ AKLT state. Using exact 
diagonalization, we obtain the spin-gap of the AKLT parent Hamiltonian. We also compute the fidelity susceptibility 
 \cite{alet2010}
of this model, and find that it indicates a direct AKLT-N\'eel transition. Using the fidelity susceptibility and the assumption of an $O(3)$ critical point, we identify the AKLT-N\'eel transition point in this model. By comparing the
spin correlations in the singlet ground state and in
the ground state with $S_z^{\rm tot}=1$,  which are obtained using the exact diagonalization,
we argue that a magnetic field applied to the AKLT state results in a transition to transverse N\'eel order.

\section{Second-neighbor exchange}

It has been suggested that the absence of N\'eel order in Bi$_3$Mn$_4$O$_{12}$(NO$_3$) is linked to
non-negligible further neighbor interactions.\cite{BiMnO} We therefore study a minimal Hamiltonian,
\bea
\label{Eq:Hnnexch}
H = J_1 \sum_{\la i j\ra} \bS_i \cdot \bS_j + J_2  \sum_{\la\la i j \ra\ra} \bS_i \cdot \bS_j  - B
\sum_i S_{i}^{z}
\eea
where $\la.\ra$ and $\la\la.\ra\ra$ denote nearest and next-nearest neighbor 
bonds respectively, and $B$ is a
Zeeman field. Let us begin with a classical analysis valid for $S\!=\!\infty$.
When $J_2\!\!=\!\!B\!\!=\!\!0$,
the ground state has collinear N\'eel order.
For $J_2\!\!=\!\!0$ and $B\!\!\neq \!\!0$,
the spins in the N\'eel state start off in the plane perpendicular to the
applied field and cant along the field direction until they are
fully polarized for $B \!> \!6 J_1 S$.
For $B \!<\! 6 J_1 S$, the spin components transverse to the magnetic field
have staggered N\'eel order for $J_2\! < \! J_1/6$; for $J_2\!>\!J_1/6$, this gives way to a one-parameter family of
degenerate (canted) spirals.\cite{mulder2010}

\begin{figure}[tb]
\includegraphics[width=3.3in]{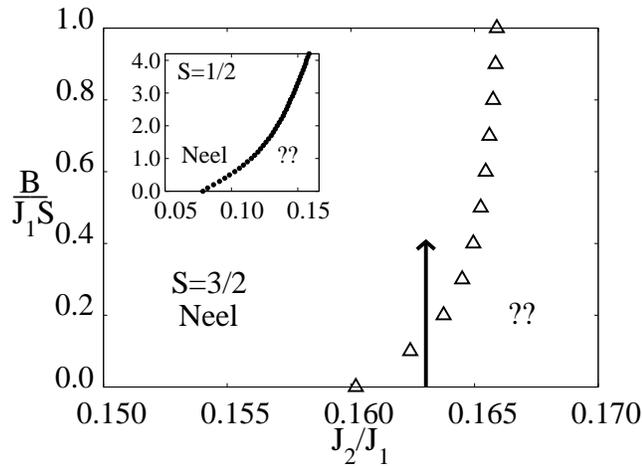}
\caption{$T\!\!=\!\!0$ melting of N\'eel order for $S\!\!\!=\!\!\!3/2$ in the 
$B$-$J_2$ plane
(open triangles) obtained using a Lindemann-like criterion, $\sqrt{\la S_x^2\ra \!-\! \la S_x\ra^2} \!\! = \!\!  5 \la S_x\ra$. 
The region ``??'' is a quantum disordered state -
possibly a valence bond solid or a quantum spin liquid.
Arrow depicts path along which one obtains
a field-induced transition to N\'eel order. (Inset) A similar melting curve for $S\!=\!1/2$.}
\label{fig:0phaseboundary}
\end{figure}

Incorporating quantum fluctuations is likely to lead to melting of N\'eel order even for $J_2 \! < \! J_1/6$.
Such fluctuations are also likely to completely suppress the classical spiral order.\cite{mulder2010}
%leading to interesting quantum paramagnets.
Using spin wave theory, we argue here that a small nonzero $B$ enhances the stability of the
N\'eel order compared to the zero field case. (i) For small nonzero $B$,
spin canting leads to a small decrease, $\propto B^2$, in the classical
staggered magnetization transverse to the field.
(ii) On the other hand,
one of the two magnon modes (labelled $\Omega_\bk^{+}$) 
acquires a nonzero gap $\propto B$ at the $\Gamma$-point
as shown in Fig.~\ref{fig:0phaseboundary}.
%(for $S\!=\!3/2$ with $J_2\!=\!0.15 J_1$ and $B\!=\!0.5 J_1 S$).  
This
suppresses low-lying spin wave fluctuations.  For $B \!\!\ll\!\! 6 J_1 S$, the latter effect
overwhelms the former, leading to enhanced stability of N\'eel order. 

Let us discuss this stability line within spin wave theory.
When $J_2 < J_1/6$ and $B < 6J_1 S$, N\'eel ordering is in the plane perpendicular to the magnetic field, but
the spins also uniformly cant in the direction of applied field, to maximally gain Zeeman energy.
The classical spin state can thus be characterized by $\bS_\br \!=\! S (\pm \cos\chi,0,\sin\chi)$ on the two sublattices.
We now define new spin operators, denoted by ${\mathbf T}_{i,\alpha}$, via a sublattice-dependent local spin rotation 
\bea
\left(\begin{array}{c}
       T_{i,\alpha}^x \\ T_{i,\alpha}^y \\ T_{i,\alpha}^z
      \end{array} \right)\!\! =\!\!
\left(\begin{array}{ccc}
       \sin\chi & 0 & (-)^{\alpha+1}\cos\chi \\
	0 & 1 & 0 \\
	(-)^\alpha \cos\chi & 0 & \sin\chi \end{array}\right)\!\!
\left(\begin{array}{c}
       S_{i,\alpha}^x \\ S_{i,\alpha}^y \\ S_{i,\alpha}^z
      \end{array} \right),
\eea
where $\alpha=1,2$, is a sublattice index and $i$ sums over each unit cell.

The ground state has all spins pointing toward the new local-$S^z$ axis. To study spin wave fluctuations, we rewrite the 
$T$ operators in terms of Holstein-Primakoff bosons as follows:
\bea
\nn T_{i,\alpha}^{z}\!\! &=& \! S-b_{i,\alpha}^\dg b_{i,\alpha}, \\
\nn T_{i,\alpha}^{x}\!\! &=& \!\sqrt{\frac{S}{2}}(b_{i,\alpha}+b_{i,\alpha}^\dg), \\
\nn T_{i,\alpha}^{y}\!\! &=& \!\frac{1}{i}\sqrt{\frac{S}{2}}(b_{i,\alpha}-b_{i,\alpha}^\dg). 
\eea

\begin{figure}[tb]
\includegraphics[width=2.8in]{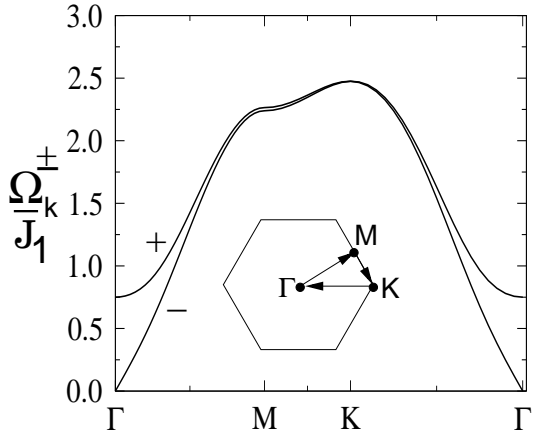}
\caption{Dispersion of magnon modes $\Omega_{\bk}^\pm$ in the $J_1$-$J_2$ model along 
depicted path in the Brillouin zone 
for $J_2\!\!=\!\!0.15 J_1$, $S\!\!=\!\!3/2$ and $B\!\!=\!\!0.5 J_1 S$. }
\vspace{0.15in}
\label{fig:magnon}
\end{figure}

The Hamiltonian can now be rewritten as $H\approx E_{Cl}+ H_{qu}$. The classical energy $E_{Cl}$ is proportional to $S^2$, and the leading order quantum correction, $H_{qu}$, is of order $S$. We get the value of the canting angle $\chi$ by demanding that terms of order $S^{3/2}$, which are linear in the boson operators, should vanish, which yields
\bea
\sin\!\chi = \frac{B}{6J_1 S}.
\eea
The classical energy is given by 
\bea
\frac{E_{Cl} }{N S^2}=-\frac{3}{2}J_1 \cos2\chi + \frac{3}{2}J_2 - \frac{B}{S} \sin\!\chi.
\eea
where $N$ is the number of sites in the honeycomb lattice. We take the magnetic field $B$ to be of order $S$, so that the Zeeman 
term $-BS_{i}^{z}$ is treated on the same level as the exchange terms $J_{ij} \bS_{i}\cdot\bS_{j}$.
The leading quantum correction is given by
\bea
\nn H_{qu} &=& -\frac{3NS}{2}J_1\cos2\chi+3NSJ_2 -\frac{NB}{2}\sin\chi    \\
&+&\sum_{\bk>0} \psi_{\bk}^\dg H_{\bk} \psi_{\bk},
\eea
where
\bea
\psi_{\bk} = \left(\begin{array}{c}
       b_{\bk,1} \\ b_{\bk,2} \\ b_{-\bk,1}^\dg \\ b_{-\bk,2}^\dg
      \end{array}\right);
%\left(\begin{array}{c}
%       a_{\bk}^\dg \\ b_{\bk}^\dg \\ a_{-\bk} \\ b_{-\bk}
%      \end{array}\right)^{T}
H_{\bk}=S\times\left(\begin{array}{cccc}
I_{\bk} & F_{\bk} & 0 & G_{\bk}\\
F_{\bk}^* & I_{\bk} & G_{\bk}^* & 0 \\
0 & G_{\bk} & I_{\bk} & F_{\bk} \\
G_{\bk}^* & 0 & F_{\bk}^* & I_{\bk}
      \end{array}\right)
\eea
with
\bea
\nn I_{\bk} &=& 3J_1 \cos2\chi - 6J_2  \\
\nn &+& \!\! 2J_2\{ \cos k_a + \cos k_b + \cos(k_a+k_b)\}\! + \!\frac{B}{S}\sin\chi, \\
\nn F_{\bk} \! &=& \!\!J_1 \gamma_{\bk} \sin^2 \chi  \equiv \vert F_{\bk} \vert e^{i\eta_{\bk}},\\
\nn G_{\bk}\!&=&\!\! -J_1 \gamma_{\bk} \cos^2\chi, 
\eea
where $\gamma_\bk = 1 + {\rm e}^{-i \bk\cdot\hb} + {\rm e}^{-i \bk\cdot(\ha+\hb)}$,
with unit vectors $\ha=\hat{x},\hb=-\hat{x}/2+\sqrt{3} \hat{y}/2$. 
This Hamiltonian can be diagonalized by a bosonic Bogoliubov transformation. The eigenvalues are given by 
\bea
\Omega_{\bk}^{\pm} = S\sqrt{(I_{\bk}\pm \vert F_{\bk} \vert)^2 - \vert G_{\bk}\vert^2}.
\eea
The Bogoliubov transformation matrix to rotate into the quasiparticle operators is given by 
\bea
\nn P=\left(\begin{array}{cc}
U_{2\times2} & 0 \\
0 & U_{2\times2}
\end{array}\right)
\left(\begin{array}{cc}
C_{2\times2} & S_{2\times2} \\
S_{2\times2} & C_{2\times2}
\end{array}\right),
\eea
where 
\bea
U_{2\times2}=\frac{1}{\sqrt{2}}\left(\begin{array}{cc}
-e^{i\eta_{\bk}} & e^{i\eta_{\bk}}\\
1 & 1 
%-\frac{F_{\bk}}{\sqrt{2}\vert F_{\bk}\vert} & \frac{F_{\bk}}{\sqrt{2}\vert F_{\bk}\vert} \\
%\frac{1}{\sqrt{2}} & \frac{1}{\sqrt{2}} 
\end{array}\right);
\eea
\bea
C_{2\times2}\!\!=\!\!\left(\begin{array}{cc}
\cosh\theta & 0 \\
0 & \cosh\phi
\end{array}\right)\!;\!
S_{2\times2}\!\!=\!\!\left(\begin{array}{cc}
\sinh\theta & 0 \\
0 & \sinh\phi
\end{array}\right),
\eea
where the angles $\theta$ and $\phi$ are given by
\bea
\nn \tanh2\theta = \frac{\vert G_{\bk} \vert}{I_{\bk}-\vert F_{\bk}\vert}, \\
\nn \tanh2\phi = \frac{-\vert G_{\bk}\vert}{I_{\bk}+\vert F_{\bk}\vert}. \\
\eea
The matrix P preserves the commutation relations of the bosonic operators and diagonalizes the Hamiltonian, giving $P^\dg H P = {\rm Diag}\{\Omega_{\bk}^{-},\Omega_{\bk}^{+},\Omega_{\bk}^{-},\Omega_{\bk}^{+} \}$.
Fig.\ref{fig:magnon} shows 
a plot of the magnon dispersion in the N\'eel state at nonzero $B$ along certain high symmetry directions in the
Brillouin zone.

The strength of long range magnetic order can be calculated in this new basis. 
For example, the in-plane component of the spin is given by
\bea
\nn\frac{1}{N}\sum_i \la S_{i,\alpha=1}^{x}\ra\!\! =\!\! (S+1/2)\cos\chi\!-\frac{\cos\chi}{N}\times \\
\sum_{\bk>0}\!\left[
 \cosh\!2\theta \{1\!\!+\!2n_B (\Omega_{\bk,-}\!)\}\!
\!+\!\cosh\!2\phi \{1\!\!+\!2n_B\! (\Omega_{\bk,+}\!)\}\right]\!,
\eea
where $n_B(.)$ denotes the Bose distribution function. For $T=0$, we can simply use the 2D Hamiltonian to 
compute this renormalized order parameter. For $T\neq 0$, we have to take into account a small coupling 
along the third dimension to allow for a stable magnetically ordered state. For a layered system with
very weak interlayer coupling, we can use the 2D Hamiltonian together with an infrared cutoff $\Lambda$ which is of
the order of the interlayer coupling. In this case, spin wave modes with energies greater than $\Lambda$ appear
to be 2D spin waves. On the other hand, modes with energies below $\Lambda$ can be dropped
as their contribution will be suppressed by phase space factors in the 3D problem.
In our numerics, we impose this infrared cutoff by simply restricting ourselves to a finite system size. 
Finite size automatically cuts off long wavelength modes with $k< k_c\sim 2\pi / \sqrt{N}$. 
In our calculations, we have restricted our system size to $2\times120\times120$ spins. This corresponds to $k_c\sim 0.05$, leading to an infrared cutoff of $\Lambda\sim0.04JS$.
%
%imposing an infrared cutoff $\Lambda$
%in the above expression 
%which is of the order the interplane coupling; modes with %energy greater than $\Lambda$ look like 2d spin waves
%and are accounted in the sum, while modes with energy lower than $\Lambda$ are dropped. We assume,

As $J_2$ is increased from zero, fluctuations around the N\'eel state increase due to frustration.
With increasing fluctuations, we expect the N\'eel state to melt when fluctuations become comparable 
to the magnitude of the ordered moment. 
To estimate the `melting curve', we assume that the transverse spin components have N\'eel order
along the $S_x$-direction, and 
use a heuristic Lindemann-like criterion for melting:
$\sqrt{\la S_x^2\ra - \la S_x\ra^2} \! > \! \alpha \la S_x \ra$. The expectation values are evaluated
(using linear spin wave theory) to order 1, even though the Hamiltonian has terms upto order $S$ only. 

\begin{figure}[tb]
\includegraphics[width=3.3in]{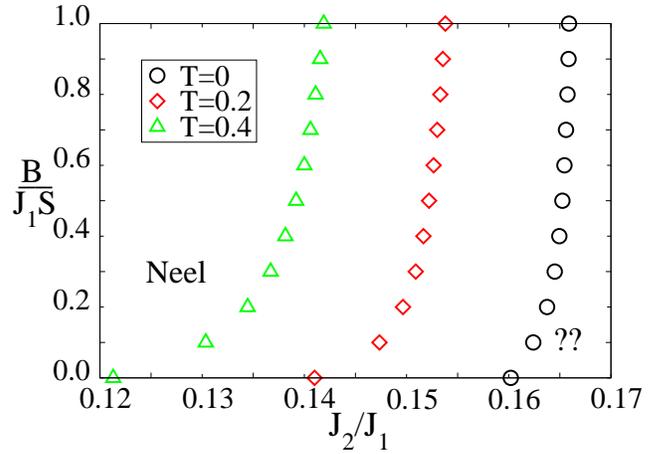}
\caption{(Color online)
Melting of N\'eel order for $S\!\!=\!3/2$ in the 
$B$-$J_2$ plane for depicted nonzero temperatures. 
To the left of the curve, there is stable canted N\'eel order. 
To the right, combined effects of quantum and thermal fluctuations melt the in-plane N\'eel order.}
\label{fig:Tphaseboundary}
\end{figure}

We set $\alpha\!\!=\!\!5$ since this leads to
melting of N\'eel order for $S\!\!=\!\!1/2$ at $J_2 \approx 0.08 J_1$, in agreement
with a recent variational Monte Carlo study by Clark {\it et al}. \cite{fawang2010}
The resulting N\'eel melting curves, at zero and nonzero temperatures, are shown in 
Fig.\ref{fig:0phaseboundary} and Fig.\ref{fig:Tphaseboundary}.

As shown in Fig.\ref{fig:0phaseboundary} and its inset, quantum fluctuations at $B=0$ 
lead to melting of N\'eel order even for $J_2 \!<\! J_1/6$ (i.e., before the classical destruction of 
N\'eel order).
For nonzero $B$, the `melting point'
moves toward larger $J_2$, leading to a window of $J_2$
over which the quantum disordered liquid can undergo a field-induced
phase transition to N\'eel order. 
The window of $J_2$ where such physics is operative appears to be small for $S\!\!=\!\!3/2$; however, 
disorder effects, which tend to suppress the stiffness, \cite{disorderboson}
may enhance this regime. 
Furthermore, as seen from Fig.~\ref{fig:Tphaseboundary}, the window of $J_2$ over which we expect field 
induced N\'eel order is also enhanced at small nonzero temperatures.
Finally, we expect field induced N\'eel order even for $S\!\!=\!\!1/2$ 
(see inset to Fig.~\ref{fig:0phaseboundary}). 

Our results are
consistent with recent neutron diffraction experiments \cite{unpub} on Bi$_3$Mn$_4$O$_{12}$(NO$_3$)
which find field induced N\'eel order.
Our results also explain recent Monte Carlo simulations
of the classical $J_1$-$J_2$ 
model  \cite{kawamura2010}  with $B\!\!\neq\!\! 0$; if $J_2\!\! = \!\!0.175 J_1$, as in the simulations,
increasing $B$ at a fixed temperature takes us closer to the melting curve as seen from Fig.~\ref{fig:Tphaseboundary}. This 
may lead to the numerically
observed enhanced N\'eel correlations. Nevertheless, we expect that there will be no field-induced 
{\it long-range} N\'eel order for $J_2\!\! = \!\!0.175 J_1$ in the classical model. 
Finally, the J$_1-$J$_2$ model is a reasonable effective model of the insulating phase
of the repulsive honeycomb lattice Hubbard model, and recent
quantum Monte Carlo simulations find a paramagnetic (spin
liquid) insulator over a range of repulsion strengths in this model. \cite{assaad}
Our prediction of field induced Ne´el order in this paramagnet can
be verified by including a magnetic field in these quantum Monte Carlo simulations.

%The next-neighbor exchange model may thus provide a plausible explanation for the experimental
%data on Bi$_3$Mn$_4$O$_{12}$(NO$_3$). We next turn to other interesting Hamiltonians which support
%quantum paramagnetic ground states that are also of broader interest.

\section{Bilayer honeycomb lattice and the interlayer dimer state} 

The Mn sites in a unit cell of Bi$_3$Mn$_4$O$_{12}$(NO$_3$) form an AA
stacked bilayer honeycomb lattice. A recent density functional theory calculation\cite{vandenBrink} estimates 
that the interlayer coupling within each bilayer is large, and may play an important role
in determining the ground state. If the interplane antiferromagnetic
exchange $J_c$ is indeed large compared to $J_1$, adjacent
spins on the two layers could dimerize, leading to loss of N\'eel order.
To study this interlayer VBS, we use the J$_1-$J$_c$ model beginning from the
limit of J$_1=0$; this leads
to the spectrum $E_j \! =\! -J_c (S (S+1)\! -\! j (j \!+\! 1)/2)$, with $j\!=\!0,1,\ldots,2S$ denoting the total spin state
of the dimer.
Restricting attention to the low energy Hilbert
space spanned by the singlet and the triplet states, we define generalized
spin-$S$ bond operators via: $|s\ra = s^\dagger|0\ra$, and $|\alpha\ra=t^\dagger_\alpha|0\ra$, where $|0\ra$ is
the vacuum, and $|\alpha(\!=\!x,y,z)\ra$
are related to the $m_j$ levels of the triplet by $|z\ra\!=\!|m_j\!=\!0\ra$, 
$|x\ra \!=\! (|m_j\!=\!-1\ra\!-\!|m_j\!=\!1\ra)/\sqrt{2}$,
and $|y\ra \!=\! i (|m_j\!=\!-1\ra\!+\!|m_j\!=\!1\ra)/\sqrt{2}$. Denoting the two spins constituting the dimer, by $\bS_\ell$,
with layer index $\ell=0/1$,
we obtain \cite{brijesh}
\bea
\bS^\alpha_{\ell} \!\!&\!\approx\!&\!\! (-1)^\ell \sqrt{\frac{S(S+1)}{3}} (s^\dagger t_\alpha^\pdg \!+\! t^\dagger_\alpha s^\pdg) \!-\! 
\frac{i}{2} \varepsilon_{\alpha\beta\gamma} t^\dagger_\beta t^\pdg_\gamma,
\eea
together with the constraint $s^\dagger s^\pdg \!+t^\dg_\alpha t^\pdg_\alpha \!=\! 1$
at each site.

To treat the effect of $J_1$, we use bond operator mean field theory \cite{sachdevbhatt1990} which
yields a reasonably accurate phase diagram for the
spin-1/2 bilayer square lattice Heisenberg model.\cite{sandvik,matsushita}
Assuming the singlets are condensed in the dimer solid,
we replace $s^\dagger\!=\!s^\pdg\!=\!\bar{s}$, and incorporate a Lagrange multiplier in the Hamiltonian
which enforces $\la t^\dg_\alpha t^\pdg_\alpha \ra = 1-\bar{s}^2$ on average. Let $N$ be
the number of spins in each honeycomb layer.
We then obtain the Hamiltonian,
\be
H = \sum_{\alpha,\bk>0} \Psi^\dagger_{\bk\alpha}  M^\pdg_\bk \Psi^\pdg_{\bk\alpha} + 2 N C, 
\ee
describing
the dynamics of the triplets. Here 
$\Psi^\dagger_{\bk\alpha}=(t^\dagger_{\bk \alpha 1} t^\dagger_{\bk \alpha 2} t^\pdg_{-\bk\alpha 1}
t^\pdg_{-\bk\alpha 2})$ (with $1,2$ denoting the two sublattices in each layer) and the 
matrix $M_\bk$ takes the form
\be
M_\bk = \begin{pmatrix} A_\bk & B_\bk & 0 & B_\bk  \\ B^*_\bk & A_\bk & B^*_\bk & 0  \\ 
0 & B_\bk & A_\bk & B_\bk \\
B^*_\bk & 0 & B^*_\bk & A_\bk \end{pmatrix}
\label{eq:Mk},
\ee
with
\bea
A_\bk &=& J_c - \mu - J_c S (S+1), \\
B_\bk &=& \frac{2}{3} \gamma_\bk J_1 S (S+1) \bar{s}^2.
\eea
Here we have defined
$\gamma_\bk = 1 + {\rm e}^{-i \bk\cdot\hb} + {\rm e}^{-i \bk\cdot(\ha+\hb)}$,
with unit vectors $\ha=\hat{x},\hb=-\hat{x}/2+\sqrt{3} \hat{y}/2$, and the constant
\be
C \!=\! - \frac{\mu}{2} (\bar{s}^2 \! -\! 1) - \frac{3}{4} (J_c \! -\!  \mu \! -\!  J_c S (S\!+\!1)) - \frac{1}{2} J_c
\bar{s}^2 S(S\!+\!1).
\ee

\begin{figure}[tb]
\includegraphics[width=3.0in]{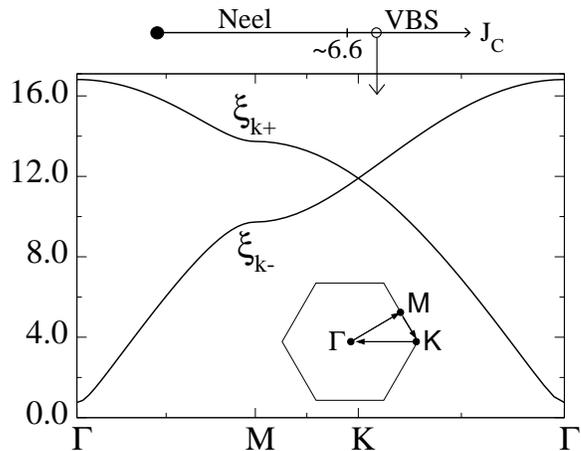}
\caption{Phase diagram of the $S\!=\!3/2$ bilayer honeycomb model obtained using
bond operator theory, and 
triplon dispersion along depicted path in the Brillouin 
zone within the interlayer
VBS state for $J_c/J_1 \!=\!7.6$ (in units where $J_1\!=\!1$).}
\label{fig:triplon}
\end{figure}

Diagonalizing this Hamiltonian leads to the 
ground state energy per spin
$ E_g \!=\! \frac{3}{2 N} \sum_{\bk > 0} (\xi_{\bk +}  + \xi_{\bk -}) \!+\! C$
where $\xi_{\bk\pm} \!=\! \sqrt{ A_\bk (A_\bk \pm 2 |B_\bk|)}$.
Setting $\partial E_g/\partial \bar{s}^2\!=\!\partial E_g/\partial\mu\!=\!0$, we obtain the 
mean field values of $\bar{s}$ and $\mu$ which minimize the ground state energy
subject to the constraint. Solving these equations numerically, we find that the spin-$S$ interlayer
VBS is a stable phase for $J_c\! > \!J_\star[S]$ where $J_\star[3/2] \! \approx \! 6.6 J_1$, $J_\star[1] \! \approx \! 3.5 J_1$
and $J_\star[1/2]\!\approx\! 1.3 J_1$.
Quantum Monte Carlo studies of this model would be valuable in firmly establishing the
value of $J_\star[S]$ as a function of $S$.
Fig.~\ref{fig:triplon} shows the triplon dispersion of the $S\!=\!3/2$ interlayer
VBS state at $J_c \!=\! 7.6 J_1$ along high symmetry cuts
in the hexagonal Brillouin zone.
%; note the low energy
%triplon mode near the $\Gamma$-point.
For $J_c \! < \! J_\star[S]$,
or in the presence of a magnetic field which can close the spin gap in the VBS state for 
$J_c \!>\! J_\star[S]$, the low energy triplon
mode at the $\Gamma$-point condenses; its eigenvector is consistent with N\'eel order.
For the field induced N\'eel state, the N\'eel ordering is
in the plane transverse to the applied magnetic field.

We see N\'eel order can be destroyed by large bilayer coupling, and the resulting 
interlayer VBS state shows field-induced N\'eel order. However, this is unlikely to be the case in Bi$_3$Mn$_4$O$_{12}$(NO$_3$) 
as it would require a very large bilayer coupling $J_c\gtrsim 6.6J_1$. This ground state could 
be realized in other honeycomb magnetic materials with lower spin. 

\section{AKLT valence bond solid} 

A particularly
interesting spin-gapped ground state of a magnet with spin-$S$ atoms on a lattice of
coordination number $z\!\!=\!\!2S$, is an AKLT valence bond state.
Each spin-$S$ is viewed as being composed of $2S$ spin-1/2
moments symmetrized on-site, with each spin-1/2 moment forming a singlet with
one neighbor. \cite{AKLT1987,arovas1988,KLT1988}
It was originally proposed as an exact realization
of Haldane's prediction of a spin-gapped ground state in 1D integer spin systems. \cite{Haldane}
Assuming that the Mn$^{4+}$ 
ions 
in Bi$_3$Mn$_4$O$_{12}$(NO$_3$) 
mainly interact with the three neighboring spins in the
same plane, this condition is satisfied with $S=3/2$ and $z=3$. The honeycomb
lattice AKLT state has
exponentially decaying spin correlations, \cite{arovas1988} and
it is the exact, and unique, zero energy ground state of the parent Hamiltonian
$
H_{\rm AKLT} = \sum_{\la ij\ra} P^{(3)}_{i,j}.
$
Here $P^{(\ell)}_{i,j}$ denotes a projector on to total spin-$\ell$ for a pair of  spins
on nearest neighbor sites $(i,j)$.
Denoting $T_{i,j}\!\equiv\! \bS_i \cdot \bS_j$, we find
\be
P^{(3)}_{i,j}= \frac{11}{128} +
\frac{243}{1440} T_{i,j}
+ \frac{116}{1440} T^2_{i,j}
+ \frac{16}{1440} T^3_{i,j}.
\ee

We have investigated, using ED (Exact Diagonalization) on clusters with
$N\!=\!12$-$18$ spins, the phase diagram of a generalized spin-3/2
model,
\be
H_Q = 
\!=\! (1-Q) \sum_{\la ij\ra} 
\bS_i\!\cdot\! \bS_j
\!+\! g Q H_{\rm AKLT},
\label{HQ}
\ee
which interpolates between a Heisenberg model (at $Q\!\!=\!\!0$) and $gH_{AKLT}$ (at $Q\!\!=\!\!1$). 
We set $g\!\!=\!\!1440/243$, so that the coefficient of 
$\bS_{i}\!\cdot\!\bS_{j}$ in $H_Q$ is unity.

\begin{figure}[tb]
\includegraphics[width=3.2in]{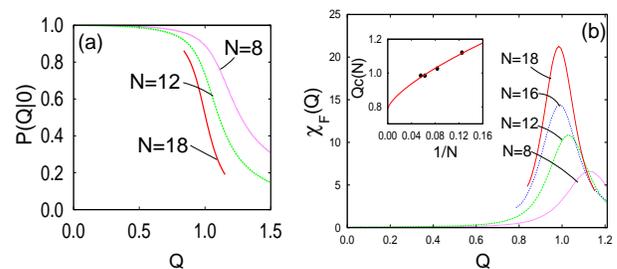}
\caption{(Color online) 
(a) Overlap $P(Q|0)$ of the ground state at $Q$ with the N\'eel state ($Q\!\!=\!\!0$) 
for various system sizes $N$, showing its rapid drop around the N\'eel-AKLT
transition. (b) Fidelity susceptibility $\chi_F(Q)$ versus $Q$ for various system sizes $N$, with
the peak indicating the N\'eel-AKLT transition point $Q_c(N)$.  Inset:
$Q_c(N)$ versus $1/N$, together with a fit $Q_c(N)\!=\!Q^\infty_c\!+\! b N^{-\frac{1}{2\nu}}$
(with a choice $\nu \approx 0.7$ assuming an $O(3)$ quantum phase transition in 2D)
which leads to $Q^\infty_c \approx 0.8$.}
\label{fig:fidelity}
\end{figure}

For $Q\!=\!0$, our analysis of the finite size spectrum shows
that the ground state energy $E_g(N,S^{\rm tot})$, as a function of
total spin $S^{\rm tot}$,
varies as $S^{\rm tot}(S^{\rm tot}\!+\!1)$,  in agreement with the expected Anderson tower
for a N\'eel ordered state. It is consistent with earlier work \cite{fouet2001,young1989,jafari2009} showing 
N\'eel order even for spin-1/2. 
To establish the N\'eel-AKLT transition as a function of $Q$, we
study overlaps $P(Q|Q')\!\! =\!\! |\la \Psi_g(Q) | \Psi_g(Q') \ra|$ of the ground state
wave functions at $Q$ and $Q'$.
As shown in Fig.\ref{fig:fidelity}(a),
the overlap $P(Q|0)$, of the
ground state wavefunction at $Q$ with the N\'eel state at $Q'\!\!=\!\!0$, is nearly unity for
$Q \!\!\lesssim\!\! 0.8$, suggesting that the ground
state in this
regime has N\'eel character.
For $0.8 \!\! \lesssim \!\! Q \!\!< \!\!1.2$,
we observe a dramatic drop of $P(Q|0)$ for all system
sizes, which indicates a N\'eel-AKLT quantum phase
transition. 

To locate the
transition more precisely,
we compute the fidelity susceptibility  \cite{alet2010}
 $\chi_F(Q) \! = 2 (\!1\!-P(Q|Q+\delta))/\delta^2$, with $\delta \! \to\! 0$,
 which measures the 
change of the wavefunction when $Q\! \rightarrow\! Q\!+\!\delta$.
Fig.\ref{fig:fidelity}(b) shows a plot of $\chi_F(Q)$ (with $\delta\!=\!0.005$). 
We observe a peak in $\chi_F(Q)$ which indicates a phase transition; this
peak shifts and grows sharper with increasing $N$. 
Assuming the thermodynamic
transition is at $Q^\infty_c$, and that the peak position $Q_c(N)$
satisfies the scaling relation 
$(Q_c(N)\!-\!Q^\infty_c) \!\!\sim \!\!N^{-1/2\nu}$, with $\nu \!\approx\!0.7$
for an $O(3)$ quantum phase transition\cite{exponent,sandvik} corresponding to triplon condensation,
we estimate $Q^\infty_c \!\! \approx \! 0.8$. Further work is necessary to confirm the nature
of the transition.
%We have also checked that the scaling of
%the peak height 
%\cite{alet2010} is consistent with $\chi_F(Q_c(N)) \sim N^{1/2\nu}$ 
%for this
%value of $\nu$.

\begin{figure}[tb]
\includegraphics[width=3.4in]{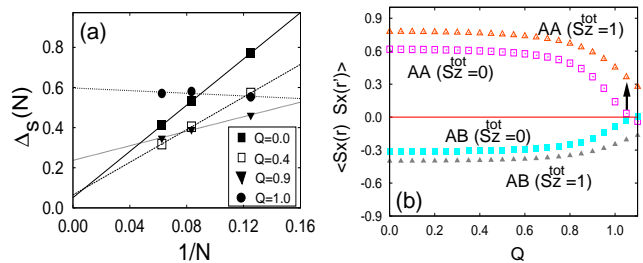}
\caption{(Color online) 
(a) Spin gap, $\Delta_s(N)$, versus $1/N$ for various $Q$,
with fits to the form $\Delta_s(N)=\Delta_s^\infty+b/N$. The small values of $\Delta_s^\infty$
for $Q\!=\!0.0,0.4$ are consistent with a gapless N\'eel state. For $Q\!=\!0.9,1.0$, the data
are consistent with a 
robust spin gap $\Delta_s^\infty$.
(b) $S_x$-spin correlations between distant sites on the same ($AA$) and
opposite ($AB$) sublattices
for $N\!\!=\!\!16$ system. The spin correlation is Neel-like ($\pm$) for $Q<1$ in the $S^{\rm tot}_z\!=\!0$ ground state; 
in the spin gapped AKLT state at Q$\gtrsim$1, it short ranged and weak in the $S^{\rm tot}_z\!=\!0$ ground state 
but it is strongly enhanced (see arrow) in the lowest lying state with $S^{\rm tot}_z\!=\!1$.}
\label{fig:corr}
\end{figure}

The spin gap of $H_Q$, $\Delta_s(N)
\!=\! E_g(N,S^{\rm tot}\!=\!1)\!-\!E_g(N,S^{\rm tot}\!=\!0)$, is plotted in
Fig.\ref{fig:corr}(a) for various $Q$ as a function
of $1/N$. Assuming a finite size scaling form
$\Delta_s(N) = \Delta_s^\infty + b/N$, we find a small value for $\Delta_s^\infty$ 
for $Q=0.0,0.4$, consistent with a gapless N\'eel state, while for $Q=0.9,1.0$
we see a robust spin gap as $1/N \to 0$.
At the AKLT
point ($Q\!\!=\!\!1$),  we estimate $\Delta^\infty_s \approx 0.6$. Given our scaling factor $g$,
this yields a value of $\approx 0.1$ for the spin gap of $H_{\rm AKLT}$.

Since the spin gap is finite for $Q\! >\! Q_c$, we expect that applying a critical field $B_{c}
\! \propto\! \Delta_s$ will lead to a phase transition; the correlation functions of
the lowest lying $S_z^{\rm tot}\!\!=\!\!1$ state at 
zero field will then reflect the correlations of the ground state 
for $B_z \!\!>\!\! B_{c}$. We plot, in Fig.\ref{fig:corr}(b), the
spin correlations on two maximally separated sites (for $N\!=\!16$) as a function of $Q$,
and make the following observations.
(i) For $S_z^{\rm tot}=0$, the ground state also has $S^{\rm tot}=0$, and
$\la S_x(i)S_x(j)\ra\!=\!\la S_z(i)S_z(j)\ra$ due to spin rotational 
invariance. At long distance, the spin correlation is strong in the N\'eel phase,
but drops rapidly to small values upon entering the AKLT state.
(ii)  In the $S_z^{\rm tot}=1$ sector,
$\la S^z(i)S^z(j) \ra \neq \la S_x(i)S_x(j) \ra$.
Remarkably, in the lowest lying state in
this sector, as opposed to the $S^{\rm tot}\!\!=\!\!0$ ground state,
we find a strong enhancement of (only) transverse correlations $\la S_x(i)S_x(j) \ra$ 
between
distant sites in the
AKLT state;
this finite-size result  suggests that the AKLT state will undergo, beyond a critical field,
a transition into a state with
in-plane N\'eel order.

\section{Discussion}
Motivated by recent theoretical
and experimental work on honeycomb lattice paramagnetic states,
we have studied various Heisenberg models with competing interactions. The models we have
studied have quantum paramagnetic ground states that undergo
field-induced phase transitions to N\'eel order. In principle, if such competing states are thought
to occur in any material, such as in Bi$_3$Mn$_4$O$_{12}$(NO$_3$),
detailed NMR
studies of isolated nonmagnetic impurities substituted for Mn may help distinguish between 
these states. For a material with $S=3/2$ moments, the interlayer VBS would have an impurity 
induced $S\!\!=\!\!3/2$ local
moment on the neighboring site in the adjacent layer. The AKLT state would 
nucleate three $S\!\!=\!\!1/2$ moments on neighboring sites
in the same plane, while spinless impurities in
spin gapped $Z_2$ fractionalized spin liquids, \cite{fawang2010,yran2010,clark2010}
do not generically lead to local moments. 
Sharply dispersing triplet excitations expected in valence
bond solids discussed here
could be looked for using single-crystal inelastic neutron scattering; 
by contrast, a spin liquid may not possess such sharp modes. 
Specific heat
experiments in a magnetic field could test for Bose-condensation
of triplet excitations as a route to N\'eel order, which we think describes the transition
of the AKLT and the interlayer dimer paramagnets.

If the ground state of Bi$_3$Mn$_4$O$_{12}$(NO$_3$) is a 
valence bond solid, disorder and Dzyaloshinskii-Moriya couplings (permitted by the bilayer 
structure) may be responsible for the observed nonzero low temperature susceptibility. 
Finally, dimer crystals with broken symmetry could also be candidate ground states
in Bi$_3$Mn$_4$O$_{12}$(NO$_3$); if this is the case, disorder must be responsible for wiping 
out the thermal transition expected of such crystals. 
These are interesting directions for future research.

\acknowledgments

We thank
J. Alicea, M. Azuma, L. Balents, G. Baskaran, E. Berg, D.
Dalidovich, Y. B. Kim, M. Matsuda, O. Starykh, and S. V. Isakov for discussions.
This research was
supported by the Canadian NSERC (RG,YJ,AP), 
an Ontario Early Researcher Award (RG,AP), and 
U.S. NSF grants DMR-0906816 and DMR-0611562 (DNS). RG and AP acknowledge the hospitality of
the Physics Department, Indian Institute of Science, and the International Center for Theoretical
Sciences while this manuscript was being written.


\begin{thebibliography}{999}
\bibitem{reviews}
L. Balents, Nature {\bf 464}, 199 (2010);
R. Moessner and A.P. Ramirez, Phys. Today {\bf 59}, 24 (2006);
T. Giamarchi, Ch. R\"uegg, O. Tchernyshyov, Nat. Phys. {\bf 4}, 198 (2008).
\bibitem{assaad}
Z. Y. Meng, T. C. Lang, S. Wessel, F. F. Assaad, and A. Muramatsu, Nature {\bf 464}, 847 (2010).
\bibitem{fawang2010}
F. Wang, \prb{\bf 82}, 024419 (2010)
\bibitem{yran2010}
Y.-M. Lu and Y. Ran, arXiv:1007.3266 (unpublished).
\bibitem{clark2010}
B. K. Clark, D. A. Abanin, and S. L. Sondhi,
arXiv:1010.3011 (unpublished).
\bibitem{AKLT1987}
I. Affleck, T. Kennedy, E. H. Lieb, and H. Tasaki,
Phys. Rev. Lett. 59, 799 (1987); 
I. Affleck, T. Kennedy, E. H. Lieb, and H. Tasaki,
Comm. Math. Phys. {\bf 115}, 477 (1988).
\bibitem{arovas1988}
D. P. Arovas, A. Auerbach, and F. D. M. Haldane, \prl {\bf 60}, 531 (1988).
\bibitem{KLT1988}
T. Kennedy, E. H. Lieb, and H. Tasaki, J. Stat. Phys. {\bf 53}, 383 (1988).
\bibitem{tcwei}
J. Cai, A. Miyake, W. D\"ur, and H. J. Briegel,
Phys. Rev. A {\bf 82}, 052309 (2010);
T.-C. Wei, I. Affleck, R. Raussendorf, arXiv:1009.2840 (unpublished);
A. Miyake, arXiv:1009.3491 (unpublished).
\bibitem{akltoptical}
J. Lavoie,
R. Kaltenbaek, B. Zeng, S. D. Bartlett, and K. J. Resch,
Nat. Phys. {\bf 6}, 850 (2010).
\bibitem{BiMnO}
O. Smirnova, M. Azuma, N. Kumada, Y. Kusano, M. Matsuda, Y. 
Shimakawa, T. Takei, Y. Yonesaki, and N. Kinomura,
J. Am. Chem. Soc., {\bf 131}, 8313 (2009);
S. Okubo, F. Elmasry, W. Zhang, M.
Fujisawa, T. Sakurai, H. Ohta, M. Azuma, O.
A. Sumirnova, and N. Kumada,
J. Phys.: Conf. Ser. {\bf 200}, 022042 (2010).
\bibitem{fouet2001}
J. B. Fouet, P. Sindzingre, C. Lhuillier,
Eur. Phys. J. B {\bf 20}, 241 (2001).
\bibitem{mattsson1994}
A. Mattsson, P. Fr\"ojdh, and T. Einarsson, \prb {\bf 49}, 3997 (1994).
\bibitem{takano2006}
K. Takano, \prb {\bf 74}, 140402 (2006).
\bibitem{mulder2010}
A. Mulder, R. Ganesh, L. Capriotti, and A. Paramekanti,
Phys. Rev. B {\bf 81}, 214419 (2010).
\bibitem{kawamura2010}
S. Okumura, H. Kawamura, T. Okubo, and Y. Motome,
J. Phys. Soc. Jpn. {\bf 79}, 114705 (2010).
\bibitem{jafari2010}
H. Mosadeq, F. Shahbazi, and S. A. Jafari, arXiv:1007.0127 (unpublished).
\bibitem{unpub}
M. Matsuda, M. Azuma, M. Tokunaga, Y. Shimakawa, and N. Kumada,
 \prl {\bf 105}, 187201 (2010).
\bibitem{Wadati}
H. Wadati, K. Kato, Y. Wakisaka, T. Sudayama, D. G. Hawthorn, T. Z. Regier,
N. Onishi, M. Azuma, Y. Shimakawa, T. Mizokawa, A. Tanaka, and G. A. Sawatzky,
arXiv:1101.2847 (unpublished).
\bibitem{vandenBrink}
H. C. Kandpal and J. van den Brink, arXiv:1102.3330 (unpublished).
\bibitem{Haldane}
F. D. M. Haldane, Phys. Lett. {\bf 93A}, 464 (1983); F. D. M. Haldane Phys. Rev. Lett. {\bf 50}, 1153 (1983).
\bibitem{alet2010}
A. F. Albuquerque, F. Alet, C. Sire, S. Capponi, Phys. Rev. B {\bf 81}, 064418 (2010).
\bibitem{brijesh}
B. Kumar, \prb {\bf 82}, 054404 (2010).
\bibitem{sachdevbhatt1990}
S. Sachdev and R. N. Bhatt, \prb {\bf 41}, 9323 (1990).
\bibitem{disorderboson}
A. Paramekanti, N. Trivedi, and M. Randeria, \prb {\bf 57}, 11639 (1998).
\bibitem{young1989}
J. D. Reger, J. A. Riera, and A. P. Young, J. Phys. Cond. Matt.
{\bf 1}, 1855 (1989).
\bibitem{jafari2009} 
Z. Nourbakhsh, F. Shahbazi, S. A. Jafari, and G. Baskaran
J. Phys. Soc. Jpn. {\bf 78}, 054701 (2009);
H. Mosadeq, F. Shahbazi, S. A. Jafari, arXiv:1007.0127 (unpublished).
\bibitem{exponent}
P. Peczak, A.M. Ferrenberg, and D.P. Landau, Phys. Rev. B {\bf 43}, 6087 (1991).
\bibitem{sandvik}
A. W. Sandvik and D. J. Scalapino
Phys. Rev. Lett. {\bf 72}, 2777 (1994).
\bibitem{matsushita}
Y. Matsushita, M.  P. Gelfand, and C. Ishii, J. Phys. Soc. Jpn. {\bf 68}, 247 (1999).
\end{thebibliography}
\end{document}